# Intermediate Phase in Germanium-Selenide glasses: A Theoretical and Experimental Study.

F.Inam[1], G. Chen[1], D. N. Tafen[2], and D. A. Drabold[1]

[1]*Dept. of Physics and Astronomy, Ohio University, Athens, OH 45701 USA*
[2]*Physics Department, West Virginia University, Morgantown, WV 26506, USA*

## Introduction

$Ge_xSe_{1-x}$ glassy alloys have been a focus of extensive research since the mid 80's because of fascinating and often complex behavior as a function of Ge content '*x*'. Much of the research on these glasses is focused on understanding the network properties like intermediate range order (IRO) and rigidity transition. The topology of the Ge-Se network, in general, is best described as the linking between 4-fold $Ge(Se_{1/2})_4$ tetrahedra and polymeric chains of 2-fold Se forming a chemically ordered continuous random network (CRN). IRO (which is often connected to the first sharp diffraction peak (FSDP) in the static structure factor) in these alloys, is reported to increase over a wide composition range [1][2][3]. Though the atomistic origin of FSDP is still a debatable subject, there are clear indications that it originates from the Ge-Ge correlations [4]. In the early 1980s, Phillips [5][6] and Thorpe [7] introduced the idea of rigidity percolation based upon constraint theory, which describes the floppy to rigid transition in covalent networks. The transition occurs when the number of constraints per atom equals the number of degrees of freedom of the system. Considering the constraints due to bond-stretching and bond-bending forces, the rigidity percolation threshold is found to exist at $<r>$ = 2.4 [7], where $<r>$ is the mean coordination of the network. In the case of $Ge_xSe_{1-x}$, the rigidity threshold corresponds to the composition $x$ = 0.20. From Raman spectroscopy and novel calorimetry studies, Boolchand and his group [8] showed that a 'double' rigidity transition occurs in $Ge_xSe_{1-x}$. A finite composition window (0.20 ≤ x ≤ 0.25) between two transitions is referred to as an intermediate phase (IP). IP is also observed in a wide range of glassy alloys including $Si_xSe_{1-x}$ and $As_xSe_{1-x}$ among others[9]. The atomistic nature of the IP is still elusive.

Raman mode frequency variation of the corner-sharing (CS) and the edge-sharing (ES) tetrahedra provides structural evidence of the IP [9]. The Raman frequency of CS tetrahedra increases linearly below the IP window and shows a power law behavior above IP, while it saturates within the IP range [9]. Raman measurements under pressure by Wang *et. al*.[10] provide further insight into the nature of IP. Above and below the IP range, the CS Raman mode shows a threshold $P_c$, below which the variation in the mode frequency with composition is negligible. In the IP window, the mode frequency increases monotonically with the applied pressure suppressing a threshold. The absence of a pressure threshold suggests the stress-free nature of the IP. The evolution of the topology of the network through the IP range has been thoroughly studied in $Si_xSe_{1-x}$ glasses [11], whose topology is believed to be similar to that of $Ge_xSe_{1-x}$. The variation in the Raman frequencies of CS, ES and Se chain mode (CM) with Si composition $x$ provide some understanding of the evolution of the connectivity between Si tetrahedra and Se chains. It is observed that for $x > 0.19$, the CS, ES and CM modes become increasingly asymmetric and require more than one Gaussian to deconvolve the spectrum. Asymmetric nature of the signal suggests increase in the coupling among Si tetrahedra due to the decrease in Se chain lengths as x increases above 0.19. At $x = 0.19$, the CM splits into two modes. The intensity of the higher mode increases to a maximum at $x = 0.26$ (second rigidity threshold) and decreases to zero as $x$ increases to 0.33. The lower mode decreases linearly above $x = 0.19$. This peculiar behavior of CM suggests the appearance of short Se chain segments in the IP window, most probably Se dimers or trimers as suggested by the authors, which link the CS and ES tetrahedra. It is interesting to note that the total constraint count for Se dimer connecting two tetrahedra (Ge-Se-Se-Ge) is equal to 3.0, the number of degrees of freedom of the network. The presence of such structural segments may explain the disappearance of pressure threshold mentioned above. The thermal response of the $A_1$ (tetrahedral breathing mode) of CS tetrahedra and Se CM mode also shows some interesting features. Wang *et. al*.[12] have shown that below the IP window, the frequency shift rate, between 100K and 300K, for $A_1$ mode remains higher than that of CM mode. Between $x = 0.2$ and $x = 0.28$, the $A_1$ shift rate decreases and the shift rate for CM mode increases and coincides with that of $A_1$ mode. The decrease of the shift rate of $A_1$ mode in the IP window also suggests the appearance of shorter Se chain segments, which would increase the coupling between tetrahedra thus making $A_1$ mode less sensitive to the temperature.

The evolution of IRO through the IP has also been studied. Wang et. al. [2] has performed X-ray scattering (XS) measurements on a wide range of $Ge_xSe_{1-x}$ compositions. They have reported a shift in the slope of FSDP position with Ge content at $x = 0.2$. The area of FSDP however, shows a plateau in the range roughly coinciding with that of IP. They attribute these features to the diminishing of longer Se chain segments above $x = 0.2$. Change in the FSDP at $x = 0.2$ is also supported by anomalous X-ray scattering (AXS) measurements [13]. Recently, Shatnawi *et. al.* [3] have reported high energy x-ray synchrotron radiation measurements coupled with x-ray absorption fine structure on these glasses. Contrary to the features reported by Wang *et al*.[2] and others [1], they did not observe any structural signature from the IP. All the structural parameters related to IRO evolve smoothly through the IP window and no sudden change at the two rigidity thresholds is observed. Although the IRO signature of the intermediate phase is not yet clear due to these conflicting reports, one can look for other structural responses, for example variation in the local density of these glasses, which can be studied using techniques like small angle X-ray scattering (SAXS). The appearance of small Se chain segments in the IP window, as suggested by Raman measurements, can affect the local densities within the crossover lengths ξ below which the network can adopt fractal like characteristics [14]. There is not much literature available on the electronic features of the IP. Taniguchi *et al.*[15] have reported photoemission (PES) and inverse photoemission (IPES) studies on these glasses. They observed a splitting of the conduction band at the first rigidity threshold with the major peak shifting towards lower energies. Recently, Novita *et. al.*[16] have observed a slight increase in the conductivity of solid electrolyte glasses $(AgI)_x(AgPO_3)_{1-x}$ in the IP range relative to 'stressed rigid' phase. In the floppy range conductivity is found to increase linearly. We have performed X-ray absorption fine structure (XAFS) measurements on these glasses and have observed a clear shift in the absorption edge within the IP range, details of which are discussed later.

The first theoretical picture of the IP was developed by Thorpe *et. al.* [17]. They modified the floppy to rigid transition concept by introducing the idea that in the floppy network, inclusion of extra bonding constraints create rigid but 'stress-free' regions. Thus when the system goes through the first transition, network becomes rigid but still in a stress-free or isostatic state. Further inclusion of bonding constraints would result in the accumulation of stress in the network, and hence the system will go through the second transition from stress-free rigid to stressed-rigid transition. The isostatic state of the network in the IP range is what they referred to as the 'self-organized' phase, in which the network responds to the addition of bonding constraints in a way to avoid the accumulation of stress. In real systems, such a self-organized state must be driven by the free energy of the network. Simulations are performed using different empirical approaches to study the realization of self-organized state of the network. Thorpe *et al.* [17] have tried to achieve a stress free state starting from a floppy network by adding constraints (bonds) in such a way to avoid 'stressed bonds'. Micoulaut and Phillips [18] used a probabilistic approach. They constructed clusters of $Ge_xSe_{1-x}$ by using size increasing cluster approximation (SICA). The basic idea is to construct a binary clustered network of species A and B by joining the $BA_{4/2}$ $Ge(Se_{1/2})_4$ tetrahedra and $A_2$ (2-fold Se) units according to the probabilities of different types of construction. They have shown that, starting from a floppy cluster (consisting mostly of $A_2$ type units), an isostatic phase can be achieved by adding $BA_{4/2}$ units in a $A_2$- $BA_{4/2}$ like isostatic bonding. The isostatic phase obtained from this construction shows a local minimum in the free energy, suggesting an equilibrium phase with respect to cross-linking of these units. Other approaches have been proposed by a number of authors [19].

Recently we have studied the models of GeSe glasses spanning over a wide composition range including IP window, obtained from first principle MD simulations [20]. The topological analysis of these models reveals the structural features, which can be attributed to a kind of self-organization associated with the IP. Such a self organization involves the re-organization of local bonding as the network evolves from a floppy Se rich environment, consisting of long Se chains connecting Ge tetrahedra, to the rigid GeSe$_2$ network which mainly consist of CS and ES tetrahedra. In particular, the network maintains a relatively high concentration of isostatic twofold Se units having one Se neighbor and one Ge neighbor through the IP window. We also observed an electronic feature of the IP in our models. The conduction band is found to shift in a reproducible and composition-dependent way through the IP range, while the valence band didn't show any considerable change through the IP window. The shift of the conduction band resembles the shift observed in the inverse photoemission studies mentioned above. To investigate this feature we have performed more precise measurements using XAFS technique and observed a similar shift in the absorption edge.

## Realistic Computer Models of the Intermediate Phase

We have made the first realistic models of Ge-Se alloys based upon *ab initio* atomic interactions in the IP range. For simulations, the approximate *ab initio* density functional code FIREBALL developed by Sankey and co-workers [21] is used. The method has been found successful for a variety of covalently bonded systems, and most especially glassy germanium selenides [22]. A sequence of Ge$_x$Se$_{1-x}$ models with Ge concentrations varying between $x = 0.10$ and $0.33$ are generated using a quench from the melt technique. Radial distribution functions calculated from these models show a reasonable agreement with X-ray diffraction experiments. Details regarding model generation and their properties may be found elsewhere [20].

Structural analysis of these models reveals interesting features in the IP range. Topological parameters like the concentration of CS and ES tetrahedra and number of rings show saturation in the IP window, which hints at the presence of the intermediate phase in our models. As the building blocks of the Ge-Se network are (Ge(Se$_{1/2}$)$_4$) tetrahedra and Se$_2$ molecules, the evolution of the network from a-Se to GeSe$_2$ glass can be described in terms of the evolution of three types of Se neighbors (Ge$_m$Se$_{2-m}$), where $m = 0,1,2$. Concentrations of these units for all the compositions are shown in Fig. 1. In the IP range, isostatic like units $m = 1$ rise to a higher concentration compared to other units, which suggests the presence of a mixture of a-Se and g-GeSe$_2$ phases. Following the SICA approach [18], we fitted these concentrations with functions $a_m + b_m P_m$, where $P_m$ is the binomial probability of the formation of unit $m$ and $a_m$ and $b_m$ are the fitting parameters. Parameter $b_m$ can be interpreted as the measure of the randomness in the evolution of the concentration of unit $m$, where $b_m = 1.0$ means completely random or stochastic and $b_m = 0.0$ means no randomness at all. Fig. 1, shows a reasonable fitting of the concentration of the three units with the $b_m = 0.98, 0.6$ and $0.7$ for $m = 0,1$ and $2$ respectively [20]. The lower value of $b_1$ shows the decrease in the randomness of the network connectivity in the IP range, where $m = 1$ units approach the higher concentration. The non-stochastic nature of the IP is also revealed in the $^{129}$I Mössbauer spectroscopy (MS) experiments on these glasses.

Studies by Bresser *et al*.[23][24] show that the ratio $I_B/I_A$ of MS integrated intensity due to chemically disordered sites B and ordered sites A deviates from the prediction of the chemically ordered continuous random network (CRN) model.

The non-stochastic nature of the IP is the key to the understanding of experimentally observed anomalies in the structural evolution of these glasses. The extent to what these models show non-random behavior is of course limited by the short time scales involved in the MD simulations, but just having a similar non-random behavior in these models should be enough to understand, at least qualitatively, the structural and electronic signatures of the IP at the atomic level.

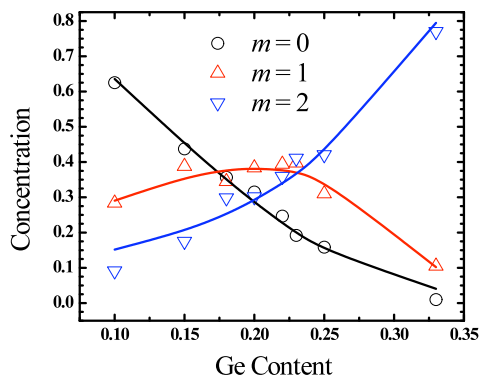

**Figure 1.** Concentration of three units (m=0,1 and 2) plotted against the Ge content. These concentrations are fitted using the probabilistic model of Micoulaut and Phillips discussed in the text [20].

## Network Evolution with Composition

The structural units of the Ge-Se network are the $Ge(Se_{1/2})_4$ tetrahedra and the polymeric Se chains. Addition of Ge would cause the cross-linking of Se chains, which then result in the appearance of relatively smaller Se chain segments $Se_n$ connecting the Ge tetrahedra, where $n = 1,2,3,…$ are the number of Se in the segment. Fig 2a shows the concentrations of $Se_n$ ($n = 2,3,4$ and $5$) segments. *All of these segments assume higher concentration in the IP window and decrease outside*. Relatively higher concentration of $n = 2$ segments for the stoichiometric composition is due to higher number of 3-fold Se sites (about 18%). Average lengths of these segments are shown in Fig. 2b. There is a subtle correlation between the concentration of these segments and their average lengths. For $n > 2$ all the chains increase in number but their lengths decrease at the first rigidity threshold, while $Se_2$ segments assume a higher concentration in the middle of the IP window, but their average lengths increase to maximum around $x = .15$ and drops to a minimum at $x = .22$, where all longer segments show increase in their average lengths. Since these segments are connected to each other through Ge sites, it is reasonable to assume some spatial correlation between bigger ($n > 2$) and smaller ($n \leq 2$) segments. Increase in the lengths of bigger segments would result in the decrease of the lengths of smaller segments. To understand the changes in the lengths and the concentrations of these segments, we focus on the behavior of $n = 1$ segments, which represent either CS or ES tetrahedra. We already have shown in Fig.1 that these segments increase from a minima in Se rich environment to a maxima at $x = .33$. Considering only the sites constituting $n=1$ segments, we calculated the average density of these sites within radial distance range 3.0 Å to 4.5 Å (which encloses the second shell in the total pair correlation function [20]) of each site (Fig. 3a). The increase in the density suggests that the Ge sites tend to form clusters rather dispersing uniformly in the a-Se network. Lets call the cluster region of $n = 1$ unit as region A, which can be considered as sub-critical segments

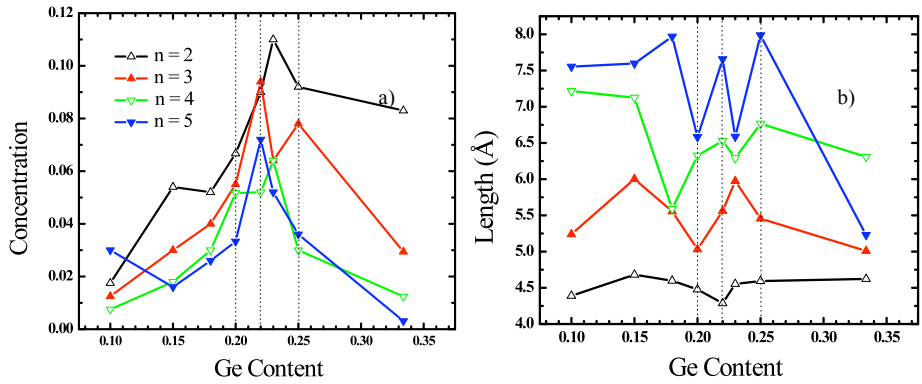

**Figure 2.** a) The concentration of short $Se_n$ (n=2,3,4,5) chains varying with the Ge content. b) the fluctuation in the average lengths of these chains.

of $GeSe_2$ embedded in the a-Se environment. Interestingly, the increase is not monotonic, density decreases in the middle of IP window before it increases again to a maxima. The decrease in the density around $x = .22$ can be explained in connection with the high concentration of longer Se segments in the middle of the IP, as shown in fig. 2a. Shorter Se segments appear due to the cross linking which occur at Ge sites. Increase in Ge content increases the cross linking and thus increases the concentration of these segments. As Ge sites form clusters of CS and ES tetrahedra, it guaranties the persistence of these Se segments. The clustering of Ge sites may be inferred from the increase in the intensity of the FSDP[3] which is mainly derived from Ge-Ge correlations [4][22]. Since the Se chains appear between these clusters, they act as a 'barrier' keeping these clusters apart. Lets call the region filled by the $Se_n$ segments with $n > 1$ region B. The rest of the space encompassing the a-Se environment, we call region C. Fig. 3b shows the variation of Se concentration which lies in the three regions. Starting from $x = .10$, Se in region A and B increases at the cost of Se sites in region C. In the IP window, the concentration of Se in region B remains nearly constant, while Se sites in region C continue to decrease with increase in the Se concentration in region A. This shows that Ge sites continue to nucleate clusters of CS/ES tetrahedra at the cost of region

30

C, until most of this region is transformed into region B. Further increase in Ge, then starts the cross-linking in region B, which results in the decrease of Se concentration above $x = .25$, and so the region A percolates over the whole space at $x = .33$. The process is further elaborated in the visual inspection of the atomic sites in the real space. Fig. 4a-h shows the network of region A (red) and region B (blue). Increase in the region A is accompanied by the increase in the B region. Between $x = .20$ and $.25$ the two regions seems to coexist without any considerable change in the overall volume of the region A. At $x = .33$ region A has spread all over the cell. From the above analysis an atomic picture of the IP emerges. In these models the intermediate phase essentially appears due to the emergence of short $Se_n$ chains, which is a consequence of the tendency of Ge sites to form clusters of CS/ES tetrahedra. There are clear indications of the appearance of these chains present in the evolution of Raman Se chain modes [11]. It seems that the evolution of these short chains defines the position and the width of the IP on the composition scale. Since these chains act as a barrier against the percolating region A, in the IP window, causing these chains to break would result in the collapse of the IP. This is what observed when iodine is introduced in $x = .25$ composition. Wang *et al*. [25] has reported the Raman and MDSC studies on the $Ge_{1/4}Se_{(3/4-y)}I_y$ alloys. The IP width is found to collapse considerably with the inclusion of iodine. Iodine being a monovalent tends to break the Se chain network thus causing the IP to collapse as suggested by the authors.

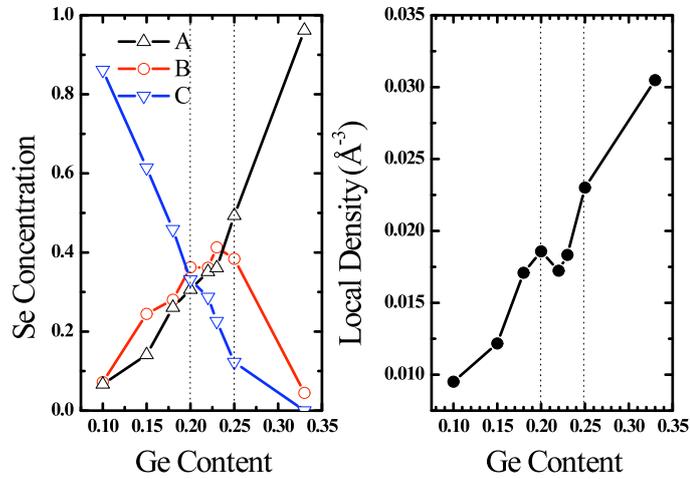

**Figure 3.** a) Concentration of Se sites in three regions A, B and C (see the text) are plotted against 'x'. b) Variation in the Local density (between the radial distance of 3.0 Å and 4.5 Å) of sites in region A is shown.

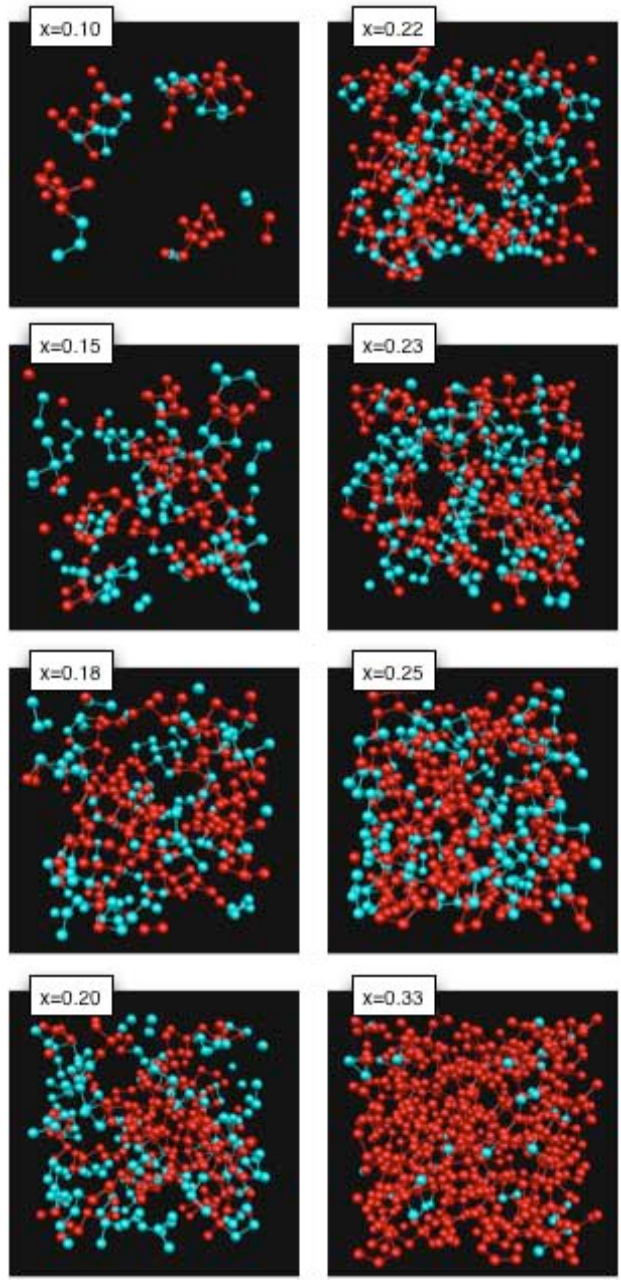

**Figure 4**. Atomic sites constituting the region A (red) [$GeSe_2$ fragments] and region B (blue) [$Se_n$ chains] are shown in real space for all the compositions.

## Electronic Structure

An earlier PES and IPES study [15] on these glasses shows some indication of the electronic consequences of the rigidity threshold. Recently we have performed X-Ray Absorption Near Edge Structure (XANES) measurements on a wide composition range of $Ge_xSe_{1-x}$ glasses including the IP window. In the following a brief description of the technique is given and the results are discussed.

**Experimental Background**

X-ray absorption fine structure (XAFS) is an experimental technique used to study atomic and electronic structure of materials. In the even of x-ray absorption by an atom, two basic processes are invoked, i.e., photon-electron interaction and electron-electron interaction. In the first process, inner shell electrons of an atom are excited to the lowest unoccupied electronic states and beyond. In the second process, the freed photoelectrons are backscattered by the neighboring atoms surrounding the absorbing atom. Interference of the outgoing electrons with the backscattered electrons modulates the x-ray absorption process. This modulation can be observed from a plot of x-ray absorption coefficient versus x-ray energy, which is called XAFS. General speaking, an XAFS spectrum is split into two regions by a point that is roughly 50 eV above the absorption edge. The lower energy region of XAFS is called XANES, and the rest part is called extended x-ray absorption fine structure (EXAFS). XANES provides information about the medium-range and the electronic structures around a specific element. In contrast, EXAFS provides complementary information about the short-range structure such as bond length, coordination number, and the Debye-Waller factor.

Structural information obtained from XANES is plenteous. On the one hand, photoelectrons in the XANES region have large wavelengths, which make XANES sensitive to medium-range structure as well as spatial correlation of short-range structure due to multiple scattering processes. On the other hand, XANES spectra contain information about the unoccupied electron density of states (viz., the conduction band) [26], Therefore, XANES can be used to study the conduction band structure.

**Experiment**

Samples selected for this study are bulk $Ge_xSe_{1-x}$ ($x$ = 0.16, 0.18, 0.19, 0.20, 0.21, 0.23, 0.24, 0.27, 0.30, 0.33) glasses prepared by the traditional melt-quenching method. Detailed description of materials fabrication can be found elsewhere[10].

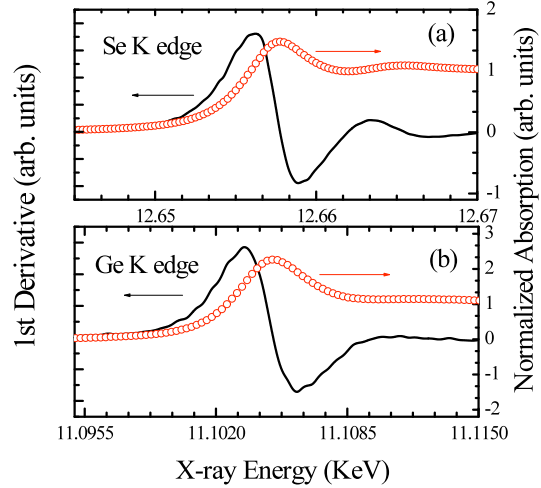

**Figure 5.** Normalized spectra (red circle) and their derivative (black line) are shown for (a) Se and (b) Ge K edge.

XANES measurements were performed on beamline 9-BM-B of the Advanced Photon Source (APS), Argonne National Laboratory. The x-ray ring at APS was operated at 7.0 GeV in non-topup mode. The circulation current of the storage ring was between 85 and 100 mA. Synchrotron x-rays originated from an undulator source were selected by a cryogenically-cooled Si (111) channel-cut monochromator. The energy resolution of the x-rays at 12 keV was about 1.5 eV. Ge K-edge ($E_{x-ray}$ = 11.104 keV) and Se K-edge ($E_{x-ray}$ = 12.658 keV) XANES spectra were collected in transmission mode. The x-ray intensities before and after the sample were recorded using two Oxford ion chambers. The samples were prepared by grinding the bulk glasses into fine powders and then spreading them onto kapton tape. Twelve layers of such tape were stacked together to optimize the signal-to-noise ratio. A $Ge_{0.33}Se_{0.67}$ powder sample was used as a reference to calibrate the absorption edge positions for all the samples. To obtain the XANES spectra with high precision, an energy step size of 0.25 eV was applied during the data collection. Figure 5 shows the normalized Ge (5a) and Se (5b) K-edge XANES spectra (red circles) of a $Ge_{0.33}Se_{0.67}$ glass. Strong x-ray absorption peaks are observed at $E_{x-ray}$ = 11.105 keV for Ge, and at $E_{x-ray}$ = 12.658 for Se. These two peaks, usually called white lines (WL), originate from the electronic transition between the 1s orbital and the conduction band. Because of the fully screened core holes in the 1s orbital, the WL is an indication of the conduction band structure [26]. Any change in the WL peak position implies a relative shift of the electronic states in the conduction band. To determine the peak position precisely, we applied the first derivative to the XANES spectra, and the results are shown in the black curves in Fig. 5. The absorption edge position was determined by finding the x-ray energy position where the first derivative equals to zero. Fig. 6 shows the variation in the relative K-edge WL positions for Se and Ge atoms with Ge content (black triangles). In the case of Se, the WL position decreases to a 'plateau' from $x =$ 0.33 as *x* decreases, with a sudden increase at *x* = .24 and *x* = .19 close to the IP window. A similar profile is observed for the WL positions for Ge atoms, though the shift is smaller compared to the shift in WL position for Se atoms. Since the XANES signal is mainly affected by local atomic arrangements around the probed site, the difference in the WL position shift for Se and Ge reflects a wider range of bonding environments among Se atoms relative to Ge atoms. Note that the Ge K-edge and Se K-edge XANES spectra were obtained independently, so the observed shift in WL positions at *x* = .19 and *x* = .24 reinforces the existence of the fluctuations in conduction edge seen in the IP window. Since we have used the same reference for all the samples, the uncertainty in the WL positions (±0.01 eV) was determined by combining the errors from repeated measurements of the same sample as well as from measurements of different samples with the same composition.

**Discussion**

The shifts in the WL positions are directly related to the shift in the conduction band. To observe this effect in our models, we calculated the average conduction band energy $<E_c> = \int_{CB} \varepsilon\, \rho(\varepsilon)\, d\varepsilon$, where $\rho(\varepsilon)$ is the electronic density of states and *CB* specifies within a quadrature range of about 1.5 eV from the conduction edge. In fig. 6, red points represent the variation of $<E_c>$ with Ge content. The absolute values are shifted with a constant for comparison. The link between models and the experiment is pleasing. The relative shift in $<E_c>$ is comparable to the shift in the WL position for Se atoms, also it shows the sudden increase at $x = .20$ and $.23$, a feature which is quite apparent in the experimental data. A small discrepancy in the positions of the two peaks from that of experiment could be due to the small size and rapid quench rates used in the preparation of these models. Apart from that, the overall profile of the shift in the absorption edge with the Ge content is clearly reproduced in the atomistic models.

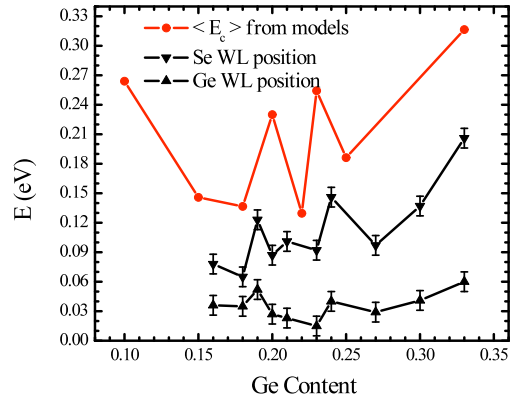

**Figure 6.** Black points shows the variation in the WL positions of Se and Ge atoms. Red points are the average conduction band edges calculated from the models.

These electronic features can be understood within the topological picture of the IP discussed above. To explore the underlying spatial character of the conduction edge states, in Fig. 7, the three lowest eigenstates at the conduction edge are shown for compositions around the IP window. Black and Red sites represent Se and Ge atoms, which contribute to the three eigenstates and white and green are their neighboring Se and Ge atoms. The shift in $<E_c>$ can be compared to the appearance of Ge neighbors (green sites) around the eigen vectors. The size of the clusters of Ge increases till x=0.18 which corresponds to the decrease in $<E_c>$ for $x > 0.10$. At $x = 0.20$ though, there are less Ge neighbors around the eigen vectors, which correlate with the increase in $<E_c>$. Similarly at $x = .22$ and $x = .25$, the Ge clusters are relatively more compact as compared to $x = .23$ compositions which shows the second peak in $<E_c>$. Since clustering of Ge sites produces short Se chains as discussed above, the region around these clusters can be considered as a strained region, which thus contribute to the localization of the eigenstates. This visual inspection implies a correlation between the clustering of Ge sites and the shift in the conduction edge energy. To explore this correlation further, we look at the evolution of atomic sites, belonging to three topological regions defined above, which contribute to the conduction and valence tail states. Fig 8a shows the concentration of atomic sites in regions A, B and C contributing to the localization of the tail states. As commented above, in the IP range the localized states are mostly localized on short $Se_n$ chains. Since in the IP range, the band tails states are mainly residing on these short Se chains (Fig. 8a), the IP would also be sensitive to the light illuminations as the micro Raman studies suggests [27]. The strain on short $n = 1$ and $n = 2$ chains, in the IP window is clear in Fig. 8b, which shows the variation in the average lengths of these chains with the Ge content. In the IP window, n=2 chains shows a clear decrease at $x = .22$ and the average length of $n = 1$ shows a profile which correlates with the shift of $<E_c>$ in the IP window. To study the effect of 'bending' of these chains on the electronic spectrum, we used a toy model consisting of $n = 1$ and $n = 2$ chains connected with each other as shown in Fig.9. First the model is relaxed by minimizing the forces using *ab initio* code VASP and the electronic density of states are calculated. To see the effect due to the bending of these chains, the chains are bent by changing angles at the Se sites keeping all the bond lengths fixed. Indeed, the bending of the chain caused the shift in the eigen spectrum with major shift in the eigenvalues above the Fermi level. The average shift above the Fermi level is about 0.8 eV, which suggest that the antibonding states of this system are strongly affected by the variation in the bond angles on Se sites.

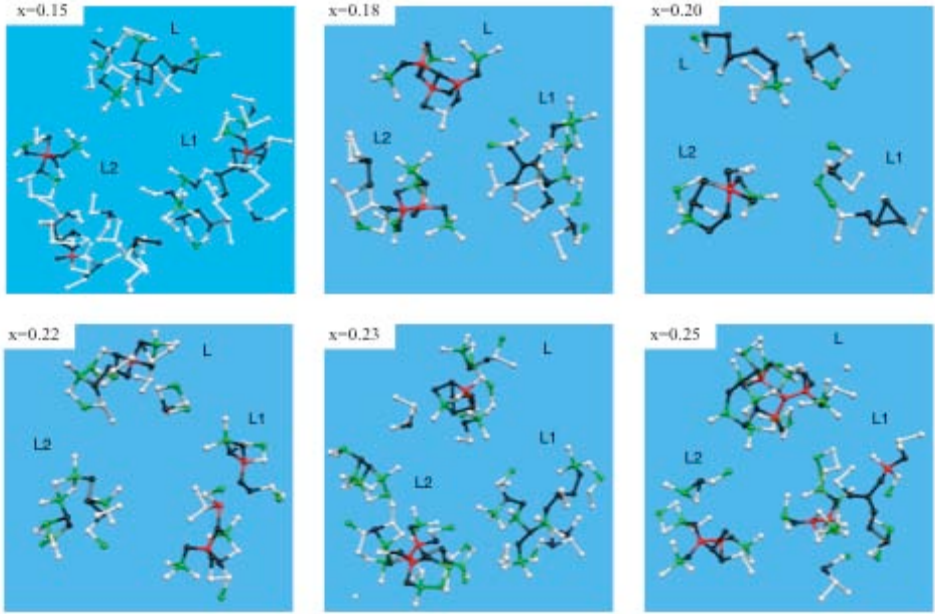

**Figure 7.** Charge from three lowest eigenstates L, L1 and L2 at the conduction edge are shown in real space for different compositions. Black and Red sites are Se and Ge contributing to the states and white and green are their neighboring Se and Ge sites.

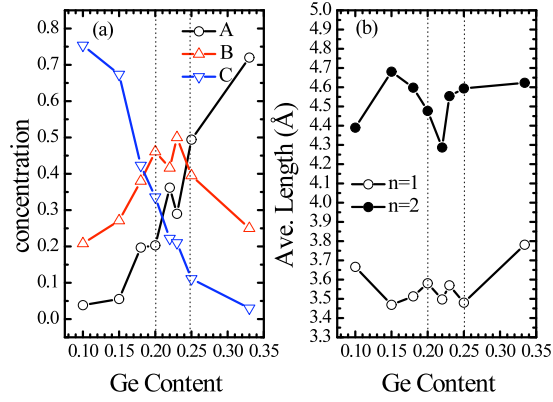

**Figure 8.** a) The variation in the atomic sites belonging to the three regions, which contribute to the localization of ten eigenstates at the valence and conduction tails. b) The variation in the average length of n=1 and n=2 Se chains.

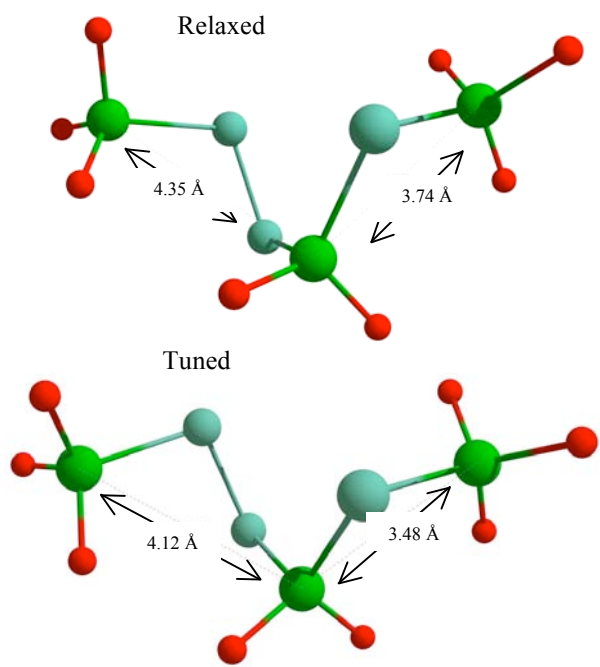

**Figure 9.** A toy model consisting of n=1 and n=2 chains. Dangling bonds on Ge (green) sites are passivated by hydrogen (red) atoms. The lengths of the chains are altered by changing angles situated on Se (blue) atoms, without changing the bond lengths and bond angles residing on Ge sites.

## Summary

From this theoretical and experimental study, a generic atomistic picture of the IP has emerged. In this picture the intermediate phase appears as a consequence of the competition between two phases (a-Se and g-GeSe$_2$). It is shown that the addition of Ge nucleates sub critical volumes of Ge clusters, which gives rise to the persisting region of short Se chains in the network. The intermediate phase appears when most of the background a-Se network is transformed to short Se chains and the system mainly consists of small volumes of CS/ES tetrahedra and short Se chains. Electronic signature of such a formation of mixed phase appears as a shift in the conduction band. This electronic feature is of the IP obtained from the models agrees well with the shift in the WL positions obtained from XANES spectra.

## Acknowledgments

We wish to strongly acknowledge Prof. P. Boolchand for providing the samples we employed in the XANES work. GC thanks Ohio University Start-up Fund for supporting this work. DAD thanks the NSF for support under Grants DMR 0605890 and DMR-0600073. FI thanks the NSF for a travel grant under DMR-0409588. Use of the APS was supported by the U. S. Department of Energy,